\documentclass[10pt,a4paper,twoside]{article}
\usepackage{./epsfig}
\usepackage{baltlat5}
\usepackage{wrapfig}

\begin{document}
\ \
\vspace{-0.5mm}

\setcounter{page}{1}
\vspace{-2mm}

\titlehead{Baltic Astronomy, vol.\ts 14, xxx--xxx, 2007.}

\titleb{Post-AGB Binaries}

\begin{authorl}
\authorb{Hans Van Winckel}{1}
\end{authorl}
\moveright-3.2mm
\vbox{
\begin{addressl}
\addressb{1}{Instituut voor Sterrenkunde, KULeuven, Celestijnenlaan
  200D, 3001 Leuven(Heverlee), Belgium}
\end{addressl}
}

\submitb{Received 2006 October 15; Revised ...}

\begin{summary}
The specific characteristic of the SED of serendipitously discovered
post-AGB binaries, allowed us to launch a very extensive
multi-wavelength study of evolved objects, selected on the basis of very
specific selection criteria. Those criteria were tuned to discover
more stars with circumstellar dusty discs. The observational study includes 
radial velocity monitoring, high spectral resolution optical studies,
infrared spectral dust studies, sub-mm bolometric observations
and high spatial resolution interferometric experiments with the VLTI.
In this contribution, we will review the preliminary results of
this program showing that the binary rate is indeed very high. We 
argue that the formation of a stable circumbinary disc 
must play a lead role in the evolution of the systems.
\end{summary}


\sectionb{1}{INTRODUCTION}

It takes a post-AGB star only about 10$^{4}$ years to evolve from a
molecular dominated AGB photosphere up to a stage where the central star
is hot enough to ionise the surrounding material.
During this short transition time, the star and circumstellar envelope must
undergo fundamental and rapid changes in structure, mass-loss mode and
geometry, which are still badly understood. The debate on which
physical mechanisms are driving the morphology changes, gained even
more impetus with the finding that also resolved but cool post-AGB stars
display a surprisingly wide variety in shapes and structure. Inspite
of intense debate between proponents
of binary models and single-star models for bipolarity and asphericity
in general (see Zijlstra these proceedings), there is still no good
understanding on the physical processes involved. One of the reasons
is that, due to the very short evolutionary phase, not many objects
are known. Very detailed studies of individual objects prevail while a
systematic study is still lacking.

The discussion also relates in a broader context to the uncertainties
involved in the models of late stellar evolution in binary systems
(see Frankowski and Podsiadlowski, these proceedings). The meeting
emphasized once more that there is, for the moment, also no clear
picture what the evolutionary relation is (if any) between the
different classes of evolved binaries like Ba stars, S-type
symbiotics, D-type symbiotics and post-AGB binaries. 

This contribution focuses on binary post-AGB stars.
The first binary post-AGB stars were serendipitously discovered
and in Sect.~2, a few well studied examples are given. This limited
sample turned out to have distinct observational characteristics 
(Van Winckel 2003) which include a broad IR excess,
often starting already at H or K. 
We started a systematic search for binary post-AGB stars based on a
sample selected by the properties of the IR-excess (Sect.~3). The preliminary
results of our ongoing radial velocity monitoring program is given in Sect.~4
and ~5. In Sect.~6 we give a small overview on the photospheric
chemical signatures observed. Sect.~7 concerns the probing of the circumbinary disc as
mayor structure element in the systems. In Sect.~8 we end with a short
discussion and highlight the mayor findings in the context of the conference.

\sectionb{2}{ILLUSTRIOUS ILLUSTRATIVE POST-AGB BINARIES}

\subsectionb{1.1}{The Red Rectangle}

One of the more famous proto-planetary nebulae is certainly the Red
Rectangle and its central star HD\,44179. The system has outstanding
properties in almost all wavelengths in which it is was studied so far
(Men'shchikov et al. 2002, Van Winckel 2003, Cohen et al. 2004 and
references therein). The central object is
not seen directly at UV and optical wavelengths, but only in the
scattered light that escapes from the polar regions of the thick disc
into the line-of-sight.

The best observational evidence that this disc must be stable comes
from the CO interferometric position-velocity maps which show that the
inner disc is indeed rotating and Keplerian speeds (Bujarrabal et al.,
2005, Bujarrabal these proceedings).  Moreover, the dust in
the inner disc is oxygen rich (Waters et al. 1998) while it is generally
accepted that the carrier of the molecular emission of the nebula (the
extended red emission) is carbon rich. The object is also one of the
strongest source of the infrared bands which are associated with
PAHs. The likeliest scenario for O-rich material in this C-rich
environment is that silicates in the long-lived circumbinary disc
antedate a recent C-rich phase of HD 44179 during which the C-rich
nebula was expelled (Waters et al. 1989).

There is now general agreement that many of unique nebular
 characteristics
of the Red Rectangle are born during phases of strong
binary interaction. The actual eccentric orbit of 318 days, is too small to
accommodate a full grown AGB star (Waelkens et al. 1996) and the object cannot have 
evolved on a single star evolutionary track.

\subsectionb{1.2}{HR\,4049}
Another serendipitously detected binary is HR\,4049, again with an
orbit which is too small to accommodate an AGB star (Waelkens et al. 1991).
The SED of this star shows a significant dust infrared excess, but with
a very peculiar temperature distribution: from J up to 850 $\mu$m, the
dust excess is consistent with a single black-body temperature of about
1150 K. While the central star of 7500K is no longer
in a dust production phase, the high dust temperature leads to the conclusion that the
dust reservoir must be very close to the central object. The detailed
modelling of the SED showed that the disc in this object must be
extremely optically thick. The significant scale height of the disc,
which is needed to account for the redistribution of stellar flux into the infrared, is a
natural consequence of the gas pressure in the inner rim of the disc
(Dominik et al. 2003).

\sectionb{3}{SYSTEMATIC SEARCH FOR POST-AGB BINARIES}

Inspired by the disc geometry of some binary objects, we launched a more
systematic search for evolved binaries with specific selection
criteria tuned to discover more objects with stable dusty reservoirs.
The total sample and selection criteria are given in
(De Ruyter et al. 2006). The IR-based selection is such that evolved objects with a hot dust component
are preferentially selected. Apart from the known binary post-AGB stars, the
sample consists of dusty RV\,Tauri pulsators and newly identified
post-AGB stars which occupy the same IRAS colour-colour region as the RV\,Tauri
stars (LLoyd Evans 1999).  Our
sample is biased towards southern objects since our radial velocity
monitoring telescope is in La Silla, Chile.

In total we selected 51 objects (De Ruyter et al. 2006),
which is a fair number compared to the
about 300 post-AGB stars known in the Galaxy (Szczerba et al, 2001).
One of the main results is that in all objects, irrespective of the
effective temperature of the central star, the dust excess starts at or
near sublimation temperature (De Ruyter et al. 2006, Fig.4 of that paper), which is
typically $<$ 10 A.U. from the central object. In most objects, the
infrared excess is a significant fraction of the total available luminosity.

\sectionb{4}{RADIAL VELOCITY PROGRAMME}

In order to be able to test our suspicion that the infrared selected
stars are binaries, we have set-up an extensive radial velocity
monitoring campaign.
Thanks to the Flemish-Swiss collaboration in exploiting the twin telescopes Mercator-Euler,
we have access to the CORALIE radial velocity spectrograph mounted on
the Swiss 1.2m Euler telescope for about 3 runs of 10 nights per
semester. We obtain spectra of limited signal-to-noise but using optimised 
cross-correlation masks, we are able to derive accurate radial velocity measurements in a
limited amount of telescope time. The line masks are tuned for every
individual star on the basis of the high signal-to-noise spectra obtained for our
photospheric chemical studies (Maas et al. 2005). We refer to the results on individual
objects for more information on the method used (Maas et al. 2002,
Maas et al. 2003).

\vbox{
\mbox{\psfig{figure=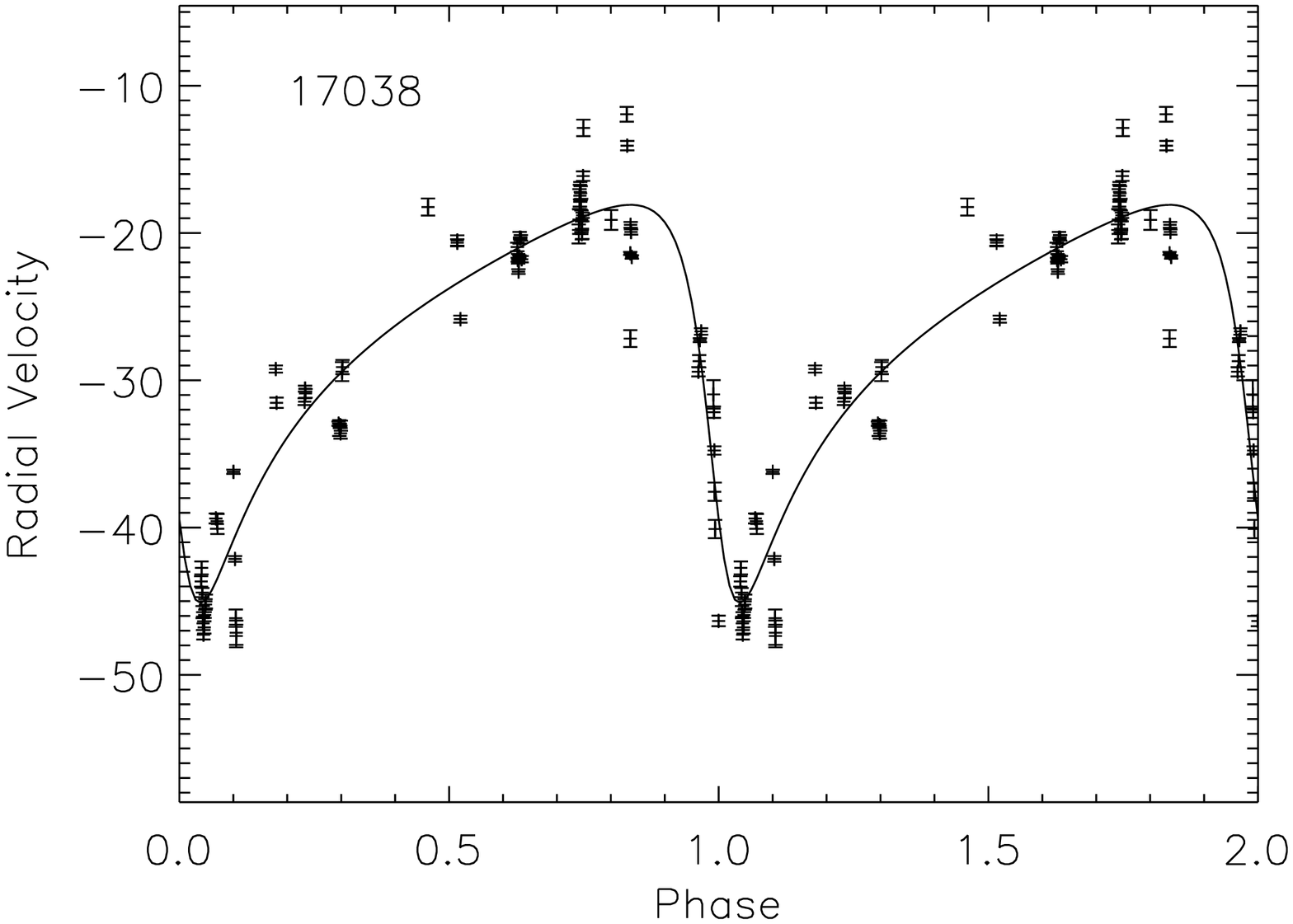,width=45truemm,angle=90,clip=}\psfig{figure=Fig17038b.ps,width=45truemm,angle=90,,bbllx=597,bblly=-10,bburx=2,bbury=839,clip=}}
\vspace{-2mm} 
\captionc{1}{Left we depict the orbital solution of the
star IRAS 17038-4815 after we cleaned the radial velocity from the
pulsation. On the right the radial velocity is given after cleaning of
the orbital motion and folded on the pulsation period. Note the strong
shock around pulsation phase 0.5. During this phase, strong
line-splitting makes the determination of the radial velocity
impossible. The orbital period is 1381 $\pm$ 16 days with a
significant eccentricity of 0.56 $\pm$ 0.05. } }
\vspace{3mm}

One of the major difficulties we encounter is that most objects are
photometric variables with significant amplitudes, also in radial
velocity. The disentangling of the photospheric motion and the
orbital motion (if present) is not straightforward and does require a
good sampling and a long monitoring baseline in time. The objects with the largest
pulsations amplitude are the RV\,Tauri stars and in Figure 1 we
show that a positive detection of the orbit is possible, despite the large
amplitude radial velocity variations due to pulsations.

Selecting only those objects with a small pulsational amplitude (of
maximal 0.25 magnitudes in the V-band), we found a binary rate of
100\%\ on six objects. The orbital periods of this sub-sample range from 120 to
1800 days with four objects showing non-circular orbits (Van Winckel
et al., in prep.). In Figure 2, two examples are given.

\vbox{
\mbox{\psfig{figure=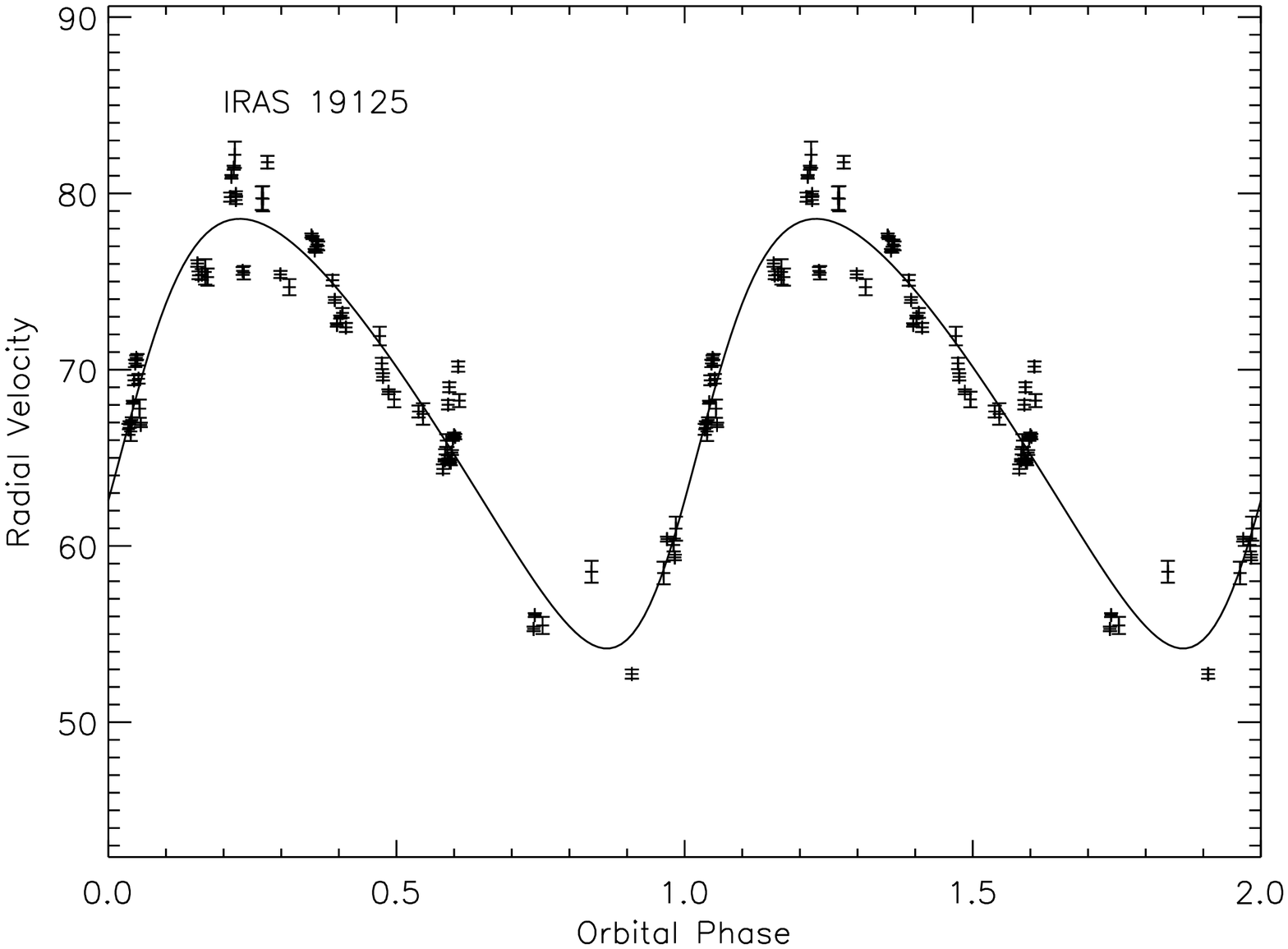,width=40truemm,angle=90,clip=}\psfig{figure=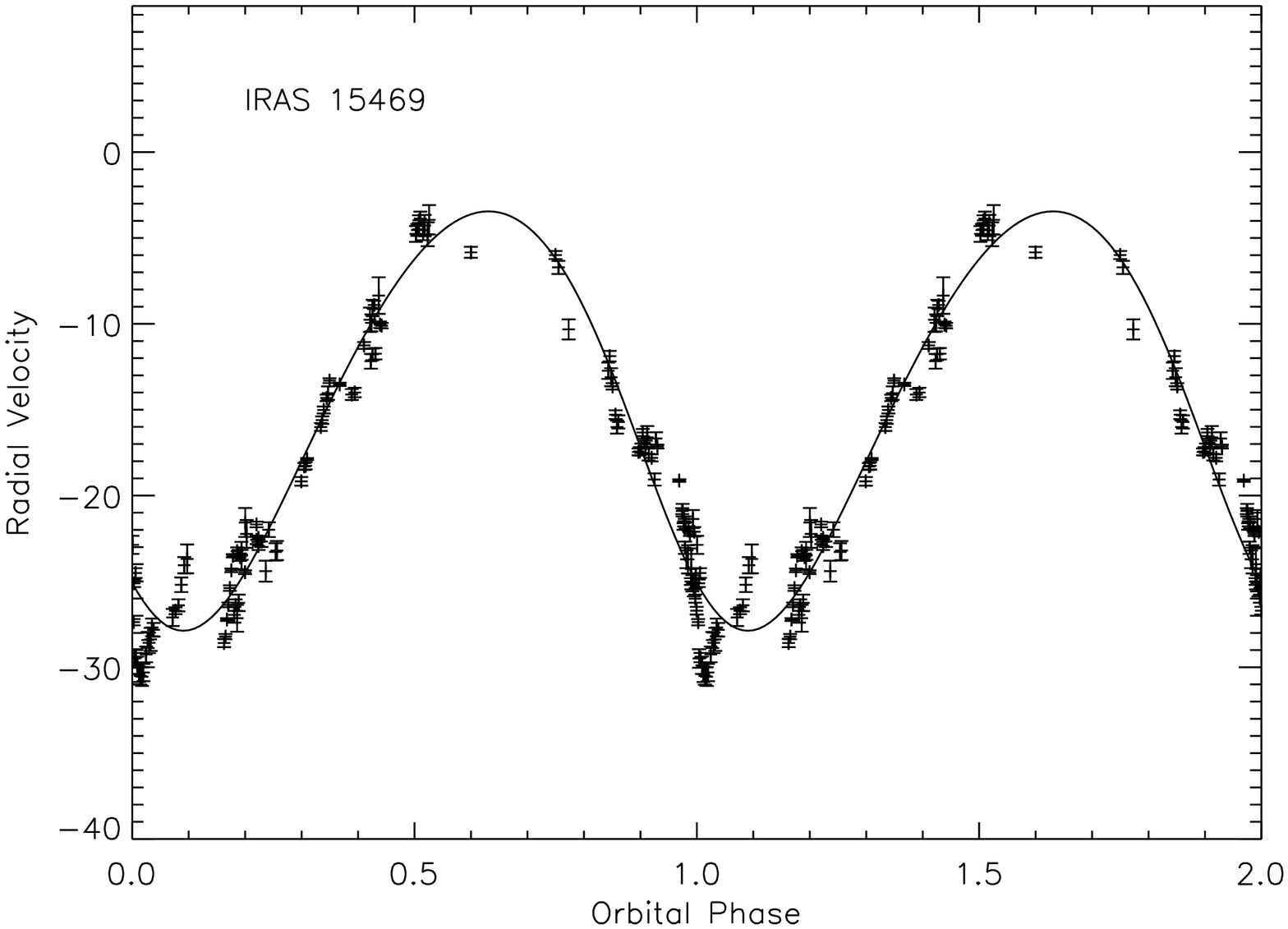,width=40truemm,angle=90,clip=}}
\vspace{-2mm} 
\captionc{2}{ Two radial velocity curves of objects in
our sample with only small amplitude photometric pulsations. The
objects are clear binaries and the radial velocities are folded on the
orbital periods.  IRAS19125$+$0343 (left panel) has a period of 517
$\pm$ 3 days and a significant eccentricity of 0.22 $\pm$
0.03. IRAS15469$-$5311 (right panel) has a period of 387 $\pm$ 1 day
and a circular orbit.  } }
\vspace{3mm}

\newpage
\sectionb{5}{ORBITAL CHARACTERISTICS OF POST-AGB BINARIES}

In Figure 3 the e-log(P) diagram is shown of all the orbits
of post-AGB stars known to date. It is an update of the figure shown
in Van Winckel (2003) which now includes all the new orbits found by our
radial velocity monitoring program. In total 28 orbits are quantified
while orbital motion of some others are suspected but no full orbit has
been sampled yet. Not the whole sample of De Ruyter et al. (2006)
is subject to our radial velocity monitoring with the Chilian
telescope due to either too faint or in the Northern Hemisphere. We
are building a spectrograph HERMES (Raskin et al. 2006) for our Mercator
telescope so also the northern sample will be subject to a detailed
monitoring study in the near future.

\vbox{
\centerline{\psfig{figure=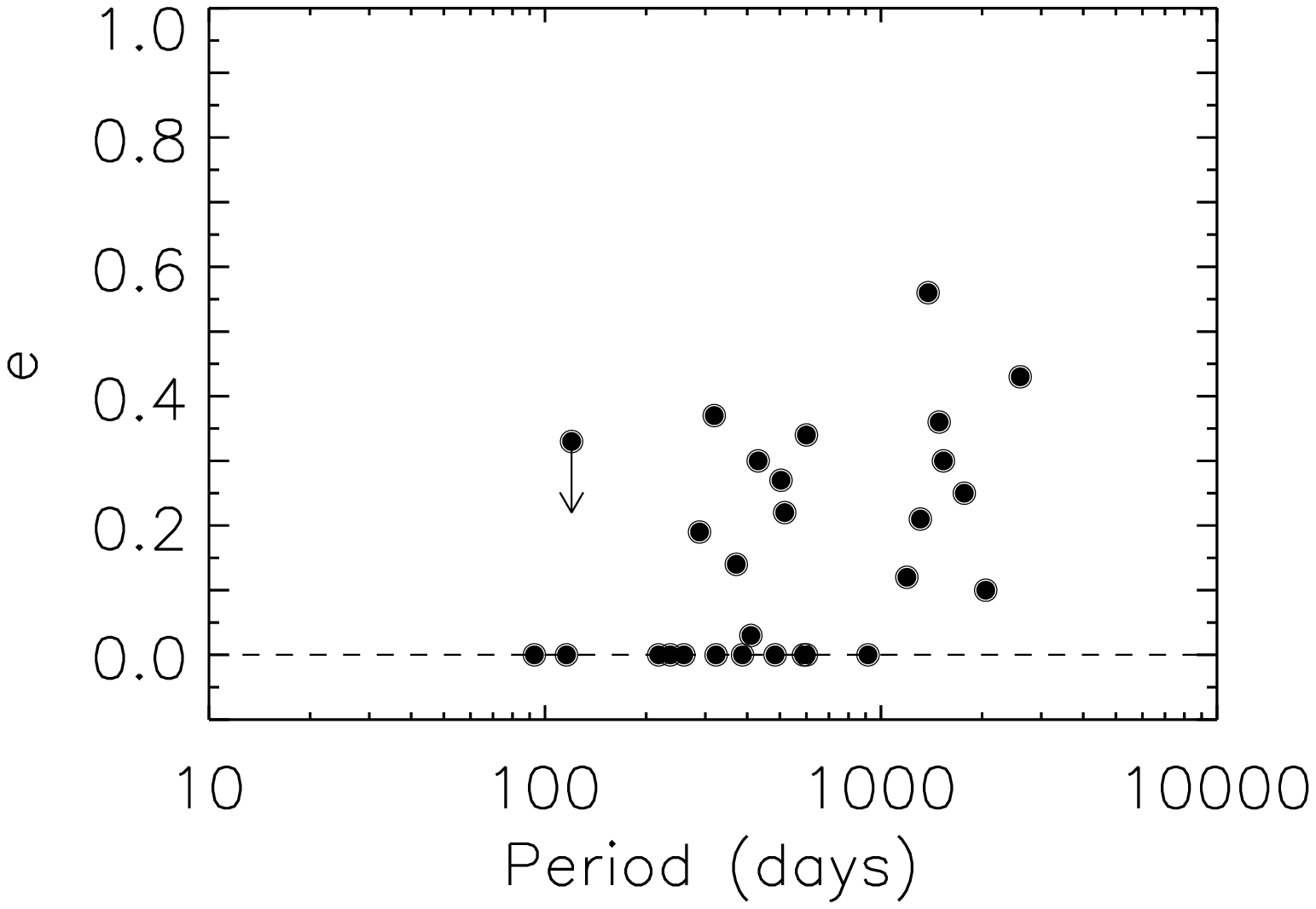,width=100truemm,angle=0,clip=}}
\vspace{-5mm}
\captionc{2}{e-log(P) diagram of the detected orbits so far. The older
  orbits (Van Winckel 2003, and references therein) are
  supplemented with the most recent results from our monitoring campaigns.}
}
\vspace{3mm}

The new orbits confirm our earlier statements (Van Winckel 2003) in 
which the excess of non-zero eccentricities is a major theoretical
challenge and still not well understood.  On general, one can say that
the actual orbits are too small to accommodate a cool AGB star with
similar luminosity. Most systems are too distant to yield significant
parallaxes and no direct tracer for their distance is available. From
the orbital distribution, it is clear that all objects must have been
subject to severe interaction when at giant dimensions. 

\sectionb{6}{THE PHOTOSPHERIC COMPOSITION}

The photospheric composition is often used as a tracer for the
internal enrichment and, especially for post-AGB stars, it could give
a very good picture of the complete AGB nucleosynthesis and dredge-up
history of an object. 

The chemical patterns which prevail in this binary sample are,
however, determined by a chemical 'depletion' process in which the separation
of the circumstellar dust from the circumstellar gas, is followed by a
selective reaccretion of only the gas, which is then rich in
non-refractory elements compared to the refractories. The photosphere
will become coated with a layer of clean gas devoid of refractory
species like Fe or Ti. Waters et al. (1992) proposed that the most
likely circumstance for the process to occur is, when the dust is
trapped in a circumstellar disc.

While this chemical depletion pattern was recognised originally in
only a few binaries (Van Winckel et al. 1995) in which the
depletion was very severe (down the [Fe/H]=$-$4.8) it turned out that
depleted photospheres are in fact rather common
(Giridhar et al. 2005, Maas et al. 2005 and references therein) and often with
only a  moderate effect on the metallicity of the object. From the 51
objects of our sample, many do show indeed depleted photospheres
(Maas et al. 2005) strengthening even more the connection between the presence of a
circumbinary disc and the depletion process.

In an oxygen-rich condensation
sequence, also the s-process elements are refractory. To trace
eventual AGB enrichment of the s-process elements one should compare the
s-process abundances with abundances of species with similar condensation
temperature. In none of the systems studied so far, there is evidence
for an overabundance of s-process elements. In a C-rich environment,
the condensation sequence is expected to be very different.

\sectionb{7}{THE CIRCUMBINARY DISC}

The compact circumstellar disc is a major structure element in all the
objects, irrespective of the orbital characteristics. The
study of the dust in the discs is a major tool to probe the
physical and chemical processes involved. Our ground based N-band spectra
(De Ruyter 2005) as well as the high resolution spectra of Spitzer
(Gielen these proceedings) illustrate the very high degree of
dust processing, both with respect to grain growth and crystallisation
of the grains. The grain growth is probed with far-infrared fluxes. 
There must be a contribution of very large ($>$\,0.1mm) grains (De
Ruyter et al. 2005) in the few objects for which we have
data so far.

Thanks to the
interferometric capabilities at ESO, we are now able to resolve the
discs and study its angular extent. Our first results (Deroo et al. 2006)
show that
the discs are indeed compact. Thanks to the spectrally dispersed
fringes of the N-band instrument MIDI, we can probe the radial
distribution of the different dust species. The hot inner rim of the
disc is above the glass temperature and grains will easily crystallise
in this environment. In the cooler outer layers, this is not the case
and cool crystalline material can be understood either as due to a
strong radial mixing within the disc, and/or a very different thermal
history of the grains at formation.
For some objects there is
evidence for a radial decrease in crystallisation but the picture is
far from clear yet (see Deroo these proceedings). 

The temperature-density distribution of a disc is
very different from the ones in outflows and calls upon detailed
radiative transfer modelling in two and even three dimensions. The
latter is especially  needed if the orbital motion of the central illuminating source is of
importance in the interpretation of the interferometric signal.
The discs are lickely very optically thick with the hot inner rim
beeing puffed up by gas pressure. In its shadow, the disc is much
cooler. Our infrared spectra show that all matrix resonances of the 
dust species show up in emission, illustrating that the surface layer of the
disc is likely optically thin at a warmer temperature that the
midplane. Detailed models of one object only was presented in
Dominik et al. (2003) but we are now in a process to model in detail the
very inner structure of the discs in many objects (see also
contributions by Baes and Vidal-P\'erez and Gielen). The model results
are confronted with the detailed analysis of the interferometric data
(see also Deroo these proceedings).

\sectionb{8}{DISCUSSION}
This contribution focuses on the post-AGB objects with hot dust in
the system. The global picture which emerges is that there are all
binary stars which are born in a system which is too small to accommodate a full grown AGB
(or in some cases maybe even an RGB) star. During a badly understood phase of strong interaction, a
circumbinary dusty disc was formed, but the binary system did not
suffer dramatic spiral in. Moreover, many systems avoided even
complete circularisation. What we observe now is a F-G supergiant in a binary system
with a dust excess starting at or near sublimation temperature. Given the
effective temperature of the central star and its high luminosity, the dust must be
circumbinary since all the determined orbits are within the sublimation radius of the dust.
The SEDs show that a considerable fraction of the luminosity of the
central star is reprocessed towards infrared radiation so the
scale height of the discs must be significant. The discs are passive
and strongly processed (see also Deroo and Gielen these proceedings).

All the discs are oxygen rich and there is no photospheric
evidence that the central star suffered from thermal pulses with
dredge-up. Given the detected orbits, the normal chemical AGB evolution was probably shortcut by a phase
of strong binary interaction. In none of the systems there is evidence
for a hot compact component and we suspect that the companion stars
are unevolved main sequence objects. 

It is interesting to note that
also in symbiotic systems, the actual white dwarf, was not s-process
and carbon enhanced when on the AGB (see Mikolajewska these proceedings). The main observational evidence
in symbiotics is that no extrinsic s-process overabundances are observed in the cool
component. On the other hand, the Ba star family (extrinsically
s-process enriched stars) show a very similar spread in orbital
periods as the one observed in post-AGB binaries, but they are clearly not the descendents
of them, because of the lack of chemical enrichment in the
evolved photosphere. For a recent review on the orbital
characteristics of the different type of systems we refer to
(Jorissen 2003). A major challenge in binary star research is
therefore to unravel the different evolutionary channels leading to
the different (chemical) classes of evolved binary stars.

We can conclude from this research that the evolved post-AGB stars with a
hot dust component are all binaries in which the dust is trapped in a
circumbinary stable disc. This disc plays a lead role in
the dynamical and chemical evolution of the system.
The formation and evolution of that dusty disc is a fundamental ingredient in the final
evolution of a significant fraction of binary stars.

\vskip5mm

ACKNOWLEDGMENTS. It is a pleasure to acknowledge the many colleagues
who are collaborating on this project: Tom Lloyd-Evans (St.-Andrews, UK),
Maarten Reyniers, Clio Gielen, Pieter Deroo, Thomas Maas (KULeuven, Belgium);
Maarten Baes, Edgardo Vidal P\'erez, Stephanie De Ruyter (UGent,
Belgium); Rens Waters, Carsten Dominik, Michiel Min (Amsterdam, The
Netherlands) as well as the many observers of the Institute of Astronomy in
Leuven for the monitoring programs.
\vskip5mm

\References

\refb
Bujarrabal V., Castro-Carrizo A., Alcolea J., \& Neri R. 2005,
 A\&A, 441, 1031

\refb
Cohen M., Van Winckel H., Bond H.~E., \& Gull T.~R. 2004, AJ, 127,
 2362

\refb
De Ruyter S. 2005, PhD thesis, UGent, Belgium

\refb
De Ruyter S., Van Winckel H., Dominik C., Waters L.~B.~F.~M., \&
 Dejonghe H. 2005, A\&A, 435, 161

\refb
De Ruyter S., Van Winckel H., Maas T., et~al. 2006, A\&A, 448, 641

\refb
Deroo P., Van Winckel H., Min M., et~al. 2006, A\&A, 450, 181

\refb
Dominik C., Dullemond C.~P., Cami J., \& Van Winckel H. 2003, A\&A,
 397, 595

\refb
Giridhar S., Lambert D.~L., Reddy B.~E., Gonzalez G., \& Yong D.
 2005, APJ, 627, 432

\refb 
Jorissen A. 2003, in {\it Asymptotic Giant Branch Stars}, ed. H.~Habing \&
 H.~Olofsson, A\&A Library, Springer, p. 461

\refb
Lloyd Evans T.~L. 1999, in {\it Asymptotic Giant Branch Stars}, eds. T. Le Bertre, A. L\`ebre, and C. Waelkens.
 IAU Symp. 1991, p. 453

\refb
Maas T., Van Winckel H., \& Lloyd Evans T. 2005, A\&A, 429, 297

\refb
Maas T., Van Winckel H., Lloyd Evans T., et~al. 2003, A\&A, 405, 271

\refb
Maas T., Van Winckel H., \& Waelkens C. 2002, A\&A, 386, 504

\refb
Men'shchikov A.~B., Schertl D., Tuthill P.~G., Weigelt G., \&
 Yungelson L.~R. 2002, A\&A, 393, 867

\refb
Raskin G., Van Winckel H., \& Lehmann H. 2006, in {\it Ground-based and
 Airborne Instrumentation for Astronomy}, Eds. by I.S. McLean, M. Iye,
 Proceedings of the SPIE, Volume 6269, p. 81

\refb
Szczerba R., G\'orny S.K., Zalfesso-Jundzi{\l}{\l}o, M., 2001, in
  {\it Post-AGB Objects as a Phase of Stellar Evolution},
  Eds. R. Szczerba and S.K. G\'orny, p. 13.

\refb
Van Winckel H. 2003, ARA\&A, 41, 391

\refb
Van Winckel H., Waelkens C., \& Waters L.~B.~F.~M. 1995, A\&A, 293,
 L25

\refb
Waelkens C., Lamers H.~J.~G.~L.~M., Waters L.~B.~F.~M., et~al. 1991,
 A\&A, 242, 433

\refb
Waelkens C., Van Winckel H., Waters L.~B.~F.~M., \& Bakker E.~J.
 1996, A\&A, 314, L17

\refb
Waters L.~B.~F.~M., Cami J., de Jong T., et~al. 1998, Nature, 391, 868

\refb
Waters L.~B.~F.~M., Trams N.~R., \& Waelkens C. 1992, A\&A, 262, L37

\end{document}